\documentclass[a4paper,11pt]{article}
\usepackage{pos}

\usepackage{siunitx}
\usepackage{subfigure}
\usepackage{wrapfig}
\usepackage{xspace}

\newcommand{\hess}{H.E.S.S.\xspace}
\newcommand{\fermilat}{\emph{Fermi}-LAT\xspace}
\newcommand{\gam}{$\gamma$\xspace}
\newcommand{\hessj}{HESS~J1809$-$193\xspace}
\newcommand{\psrj}{PSR~J1809$-$1917\xspace}
\newcommand{\psrjother}{PSR~J1811$-$1925\xspace}

\newcommand{\snrS}{G011.0$-$00.0\xspace}
\newcommand{\fermijten}{J1810.3$-$1925e\xspace}
\newcommand{\fermijeleven}{J1811.5$-$1925\xspace}

\DeclareSIUnit\erg{erg}

\makeatletter
\def\NAT@def@citea{\def\@citea{\NAT@separator}}%
\makeatother

\title{Revisiting \hessj\ --- a very-high-energy gamma-ray source in a fascinating environment}
\ShortTitle{\hessj}

\manuallySeparateAuthors
\author*[a]{L. Mohrmann}
\author[b]{, V. Joshi}
\author[a]{, J. Hinton}
\author[b]{ and S. Funk}
\author{ for the \hess Collaboration}

\affiliation[a]{Max-Planck-Institut f\"{u}r Kernphysik,\\
  Saupfercheckweg 1, 69117 Heidelberg, Germany}
\affiliation[b]{Friedrich-Alexander-Universit\"at Erlangen-N\"urnberg, Erlangen Centre for Astroparticle Physics,\\
  Erwin-Rommel-Str. 1, 91058 Erlangen, Germany}

\emailAdd{lars.mohrmann@mpi-hd.mpg.de}

\abstract{
  \hessj is one of the unidentified very-high-energy gamma-ray sources in the \hess Galactic Plane Survey (HGPS).
  It is located in a rich environment, with an energetic pulsar and associated X-ray pulsar wind nebula, several supernova remnants, and molecular clouds in the vicinity.
  Furthermore, \hessj was recently detected at energies above 56 TeV with HAWC, which makes it a PeVatron candidate, that is, a source capable of accelerating cosmic rays up to PeV energies.

  We present a new analysis of the TeV gamma-ray emission of \hessj with \hess, based on improved analysis techniques.
  We find that the emission is best described by two components with distinct morphologies and energy spectra.
  We complement this study with an analysis of Fermi-LAT data in the same region.
  Finally, taking into account further multi-wavelength data, we interpret our results both in a hadronic and leptonic framework.
}

\FullConference{%
  7th Heidelberg International Symposium on High-Energy Gamma-Ray Astronomy (Gamma2022)\\
  4-8 July 2022\\
  Barcelona, Spain\\}

\begin{document}
\maketitle

\vspace{-0.3cm}
\section{Introduction}\vspace{-0.3cm}
\hessj is an unassociated very-high-energy (VHE; $E>\SI{100}{GeV}$) \gam-ray source that was discovered in 2007 \cite{HESS_J1809_2007} as part of the \hess Galactic Plane Survey (HGPS; \cite{HESS_HGPS_2018}).
It is located close to the energetic pulsar \psrj (spin-down power $\dot{E}=\SI{1.8e36}{\erg\per\second}$, characteristic age $\tau_c=\SI{51}{kyr}$ \cite{Manchester2005}, distance $d\approx\SI{3.3}{kpc}$ \cite{Parthasarathy2019}), which powers an X-ray pulsar wind nebula (PWN; see e.g.\ \cite{Anada2010}).
Initially, \hessj was interpreted as being connected to this PWN, that is, due to inverse Compton (IC) emission from high-energy electrons accelerated in the pulsar wind (the ``leptonic scenario''; \cite{HESS_J1809_2007}).
However, the region also harbours several supernova remnants (e.g.\ \snrS, at a distance of $d\approx\SI{3}{kpc}$ \cite{Castelletti2016}, which has been proposed as the progenitor SNR of \psrj \cite{Voisin2019}, although the association is not firm) as well as dense molecular clouds \cite{Castelletti2016,Voisin2019}.
This has motivated an interpretation of \hessj in a ``hadronic scenario'', in which the \gam-ray emission is due to the interaction of cosmic-ray nuclei -- accelerated at the SNR shock front -- with gas in the molecular clouds \cite{Castelletti2016,Araya2018}.
Recently, the HAWC experiment has detected \gam-ray emission from \hessj up to energies of $\sim$\SI{100}{TeV} \cite{HAWC2020_UHE}.
Here, we present a summary of a new \hess analysis of \hessj, which is complemented by a \fermilat analysis of the same region.
For further details, we refer to the full publication about this study, which is currently under journal review \cite{HESS2022_HESSJ1809}.

\vspace{-0.3cm}
\section{Data analysis}\vspace{-0.3cm}
\hess is an array of Cherenkov telescopes sensitive to \gam rays in the \SI{100}{GeV}--\SI{100}{TeV} energy range, located in Namibia \cite{HESS2006_Crab}.
Here, we used \SI{93.2}{h} of data taken on \hessj with the four \SI{12}{m} diameter telescopes.
For the high-level analysis, we have employed the \texttt{Gammapy} package (v0.17; \cite{Deil2017,Deil2020}) and carried out a spectro-morphological likelihood analysis that uses as input a background model constructed from archival \hess observations (see \cite{Mohrmann2019} for details).
The energy threshold of the combined data set is \SI{0.27}{TeV}.

\fermilat is a pair conversion detector onboard the \emph{Fermi} satellite, sensitive to \gam rays between $\sim$\SI{20}{MeV} and $\sim$\SI{300}{GeV} \cite{FermiLAT2009}.
For the \fermilat analysis, we have used \SI{12.4}{yr} of data and employed the \texttt{Fermitools}\footnote{\url{https://fermi.gsfc.nasa.gov/ssc/data/analysis/software}} (v2.2.0) and \texttt{Fermipy}\footnote{\url{https://fermipy.readthedocs.io}} (v1.1.5) packages.
We analysed events passing the \texttt{P8R3\_SOURCE} event selection (event class~128, event type~3), using a binned analysis.

\vspace{-0.3cm}
\section{Results}
\begin{figure}[t]
  \centering
  \includegraphics[width=0.98\textwidth]{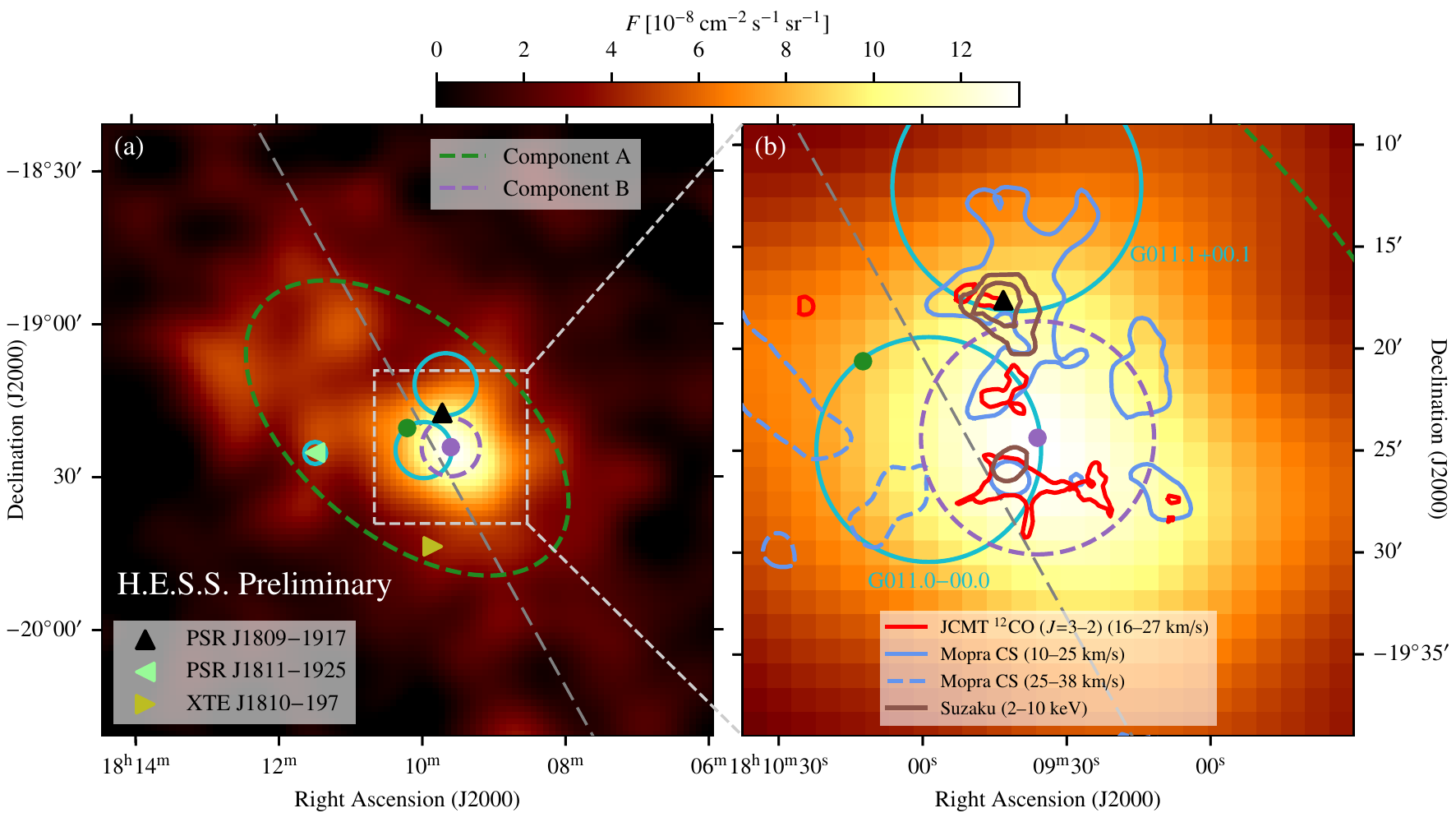}
  \caption{
    Map showing the \gam-ray flux above \SI{0.27}{TeV} from \hessj.
    (a) full region.
    (b) zoom-in on core region.
    The position of \psrj is marked with a black triangle, cyan circles denote the positions of SNRs.
    The green/purple dot and lines display the position and extent of the two components (A/B) of \hessj (cf.\ also Fig.~\ref{fig:sign_maps_hess}).
    The grey dashed line marks the Galactic plane.
  }
  \label{fig:flux_maps}
\end{figure}

\begin{figure}[b]
  \centering
  \includegraphics[width=0.98\textwidth]{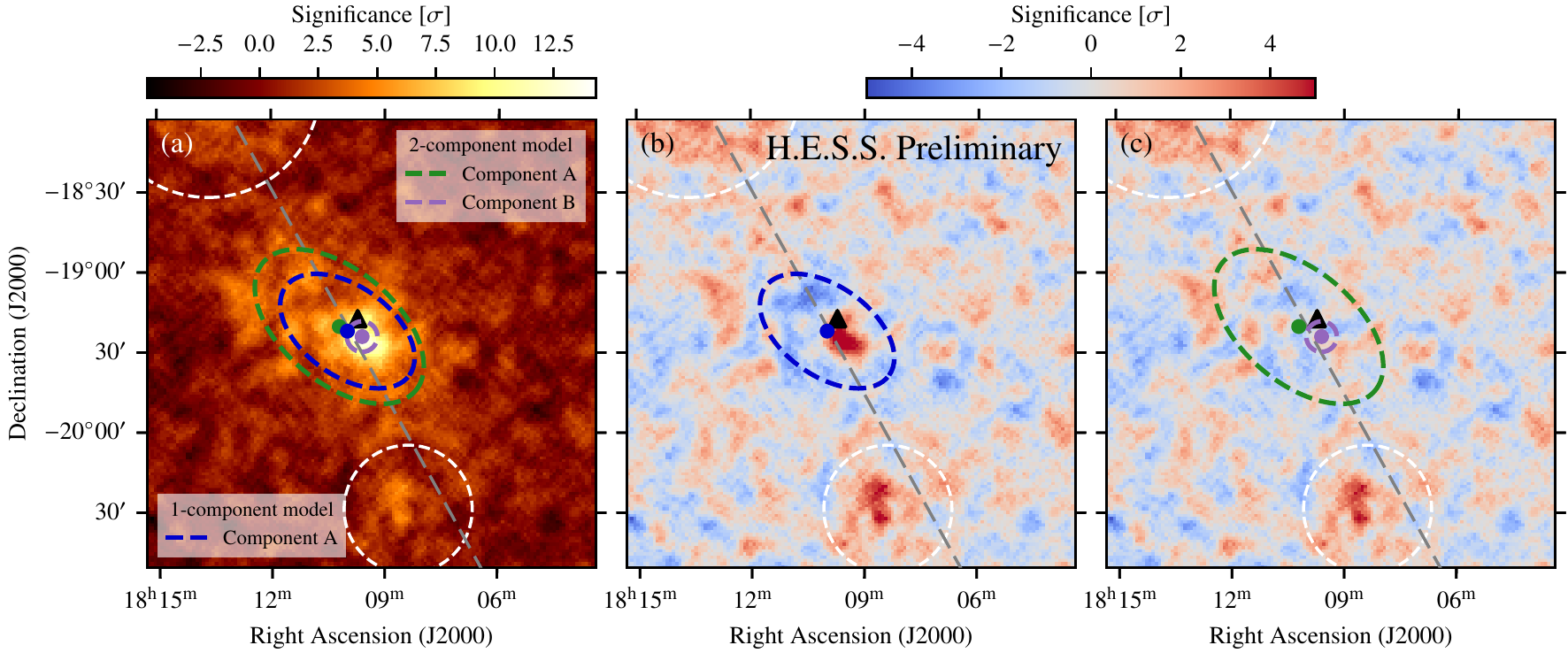}
  \caption{
    \hess significance maps for \hessj.
    Panel (a) shows the pre-modelling map, whereas panels (b) and (c) show the residual significance map for the 1-component and the 2-component model, respectively.
    White dashed circles denote regions excluded from the analysis.
  }
  \label{fig:sign_maps_hess}
\end{figure}

\vspace{-0.3cm}
In Fig.~\ref{fig:flux_maps} we show flux maps of the \hessj region.
The source is extended on a scale of about $1^\circ$, and shows a bright peak of emission close to its centre.
The significance maps in Fig.~\ref{fig:sign_maps_hess} illustrate our modelling of \hessj.
First, we have attempted to model the source with a single component that uses an elongated Gaussian as the spatial model.
However, as is evident from Fig.~\ref{fig:sign_maps_hess}(b), the model is not capable of describing the extended emission and the bright peak simultaneously.
We therefore adopted a 2-component model, in which a second component is added to describe the compact bright peak (using a symmetric Gaussian as the spatial model).
Fig.~\ref{fig:sign_maps_hess}(c) shows that this model yields a satisfactory description of the data (statistically, it is preferred by $13.3\sigma$ over the 1-component model).
We refer to the two components as component~A and~B, respectively.
Component~A has a 1-$\sigma$ major-axis extent of $\sigma_\mathrm{A}=(0.62\pm 0.03_\mathrm{stat}\pm 0.02_\mathrm{sys})\,\mathrm{deg}$ and an eccentricity of $e_\mathrm{A}=0.82\pm 0.03_\mathrm{stat}$, whereas for component~B $\sigma_\mathrm{B}=(0.095\pm 0.007_\mathrm{stat}\pm 0.003_\mathrm{sys})\,\mathrm{deg}$.

\clearpage

\begin{wrapfigure}{R}{0.54\textwidth}
  \centering
  \includegraphics[width=0.5\textwidth]{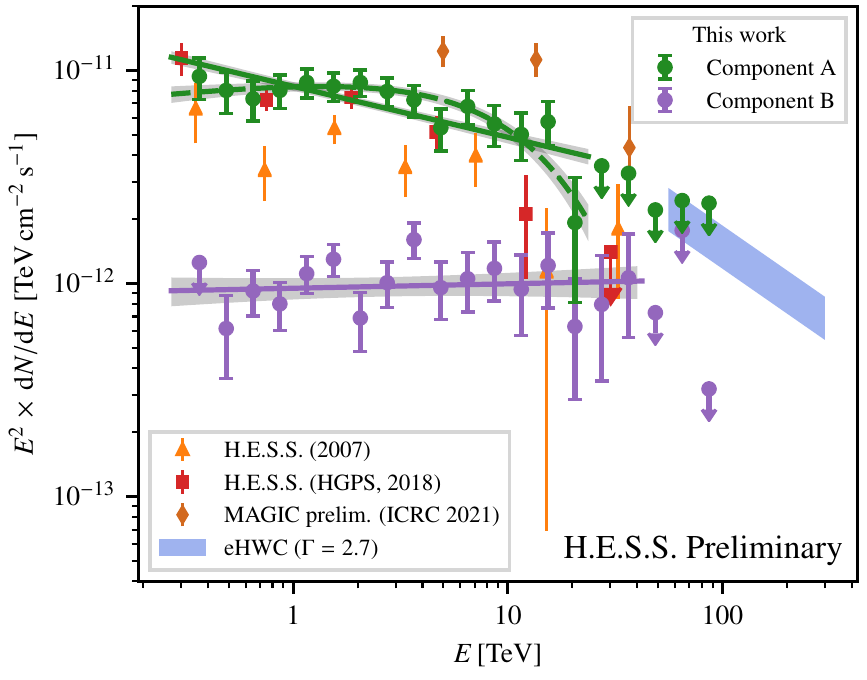}
  \caption{
    Energy spectrum of \hessj.
    The spectra of component~A and~B are shown in green and purple, respectively.
    The solid lines show the best-fit PL models for each component, and the dashed green line the best-fit ECPL model for component~A.
    Published spectra are taken from \cite{HESS_J1809_2007,HESS_HGPS_2018,Zaric2021,HAWC2020_UHE}.
  }
  \label{fig:sed}
\end{wrapfigure}

The energy spectra of the two components -- which are modelled simultaneously -- are shown in Fig.~\ref{fig:sed}.
When fitting power-law (PL) models, $\mathrm{d}N/\mathrm{d}E\propto (E/\SI{1}{TeV})^{-\Gamma}$, to both components, we obtained spectral indices of $\Gamma_\mathrm{A}=2.24\pm 0.03_\mathrm{stat}\pm 0.02_\mathrm{sys}$ and $\Gamma_\mathrm{B}=1.98\pm 0.05_\mathrm{stat}\pm 0.03_\mathrm{sys}$ for component~A and~B, respectively.
However, the upper limits at high energies for component~A indicate that the spectrum may cut off before reaching \SI{100}{TeV}.
Indeed, a power law with exponential cut-off (ECPL), $\mathrm{d}N/\mathrm{d}E\propto (E/\SI{1}{TeV})^{-\Gamma}\cdot\exp(-E/E_c)$, is preferred (by $8\sigma$) for this component, in which case we obtained a spectral index $\Gamma_\mathrm{A}=1.90\pm 0.05_\mathrm{stat}\pm 0.05_\mathrm{sys}$ and a cut-off energy of $E_c^\mathrm{A}=(12.7_{-2.1}^{+2.7}|_\mathrm{stat}\,_{-1.9}^{+2.6}|_\mathrm{sys})\,\mathrm{TeV}$.
For component~B, an ECPL model is not significantly preferred over the PL model.

\begin{figure}[b]
  \centering
  \includegraphics[width=0.98\textwidth]{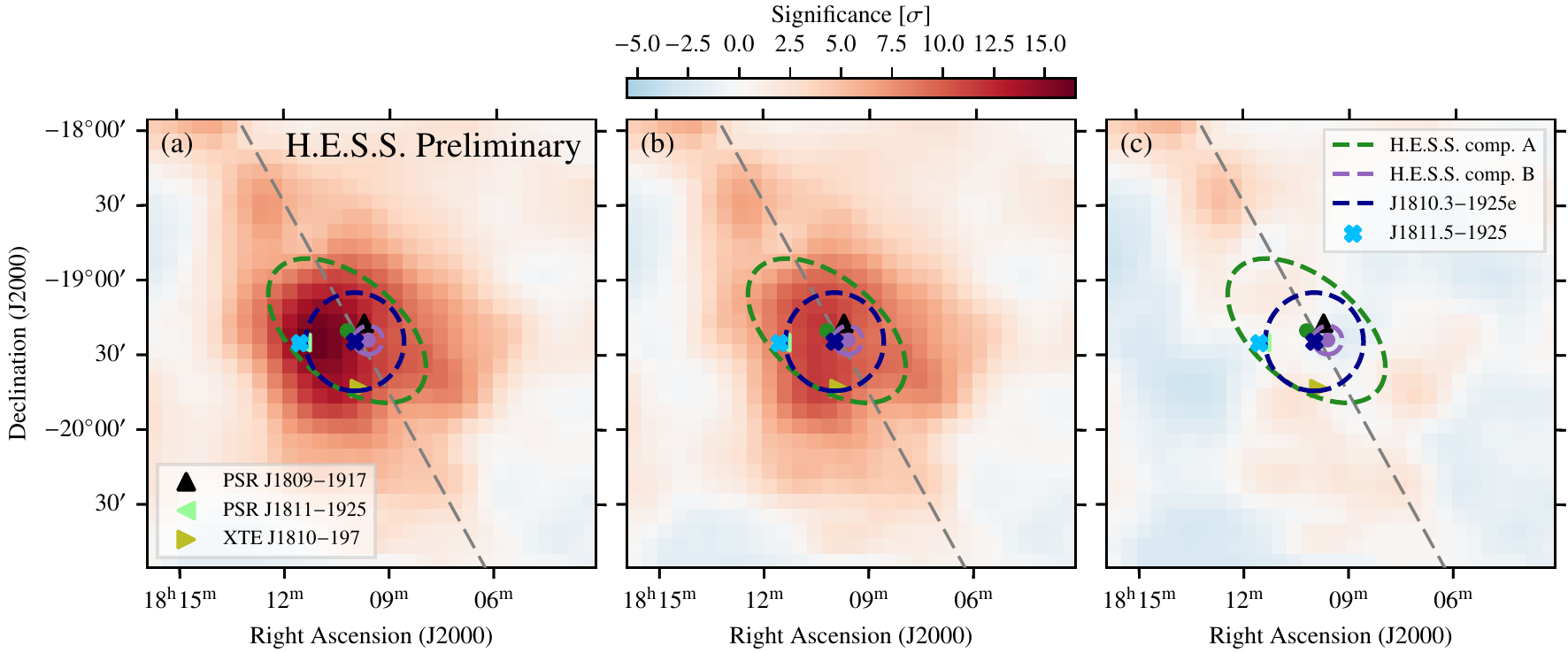}
  \caption{
    Significance maps for the \fermilat analysis.
    (a) Pre-modelling map.
    (b) With \fermijeleven in the model.
    (c) With \fermijeleven and \fermijten in the model.
    The two components of \hessj are displayed as well.
    The grey dashed line marks the Galactic plane.
  }
  \label{fig:sign_maps_fermi}
\end{figure}

In Fig.~\ref{fig:sign_maps_fermi}, we illustrate the results of the \fermilat analysis.
Similarly to the case of \hess, extended emission around \psrj is visible, although no bright peak that would correspond to component~B of \hessj can be identified.
Following the \fermilat 4FGL-DR2 catalogue \cite{FermiLAT2020,FermiLAT_4FGLDR2_2020}, we modelled the emission with two sources: \fermijeleven, which is modelled as a point source and connected to the nearby pulsar \psrjother (i.e.\ unrelated to \hessj), and \fermijten, which is modelled as an extended source.
The energy spectrum of \fermijten, exhibiting a spectral index of $\Gamma\approx 2.5\pm0.1$, is displayed in Fig.~\ref{fig:gamera}.

\vspace{-0.3cm}
\section{Discussion}\vspace{-0.3cm}
The similarity of the spatial models of component~A of \hessj and the \fermilat source \fermijten (cf.\ Fig.~\ref{fig:sign_maps_fermi}) suggests a connection between these two components.
However, the energy spectrum of \fermijten below \SI{10}{GeV} is considerably steeper than that of component~A, implying the need of a spectral break at around \SI{0.1}{TeV} if both are connected.
On the other hand, the spectrum of \fermijten could be connected to that of component~B more smoothly (although a break would still be required), but in this case its spatial extent would greatly exceed that of its counterpart.
This illustrates that a joint modelling of the emission detected with \hess and \fermilat is very challenging.
We focus here on modelling the \hess components.

\begin{figure}[b]
  \centering
  \subfigure[Full energy range.]{
    \includegraphics[width=0.48\textwidth]{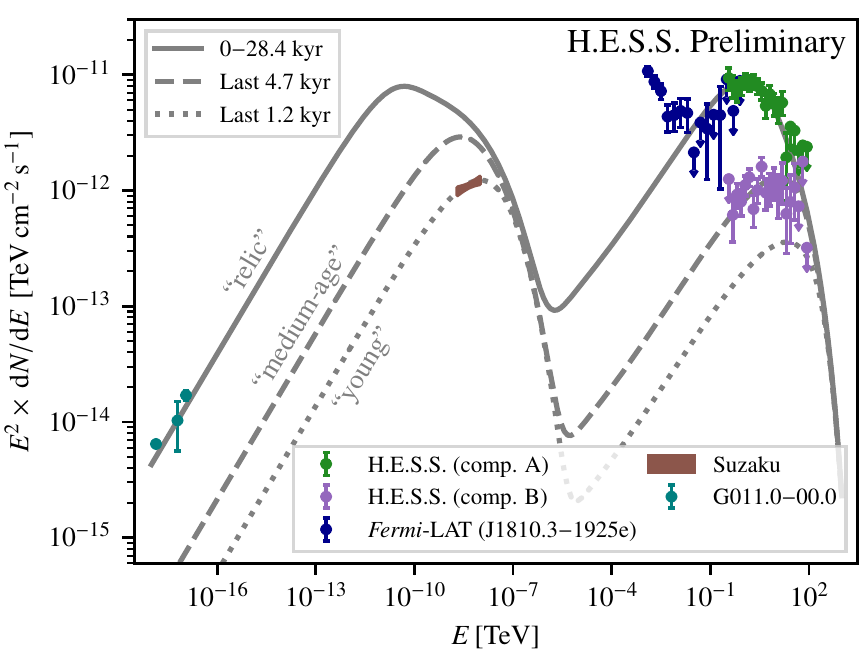}
    \label{fig:gamera_full}
  }
  \subfigure[Zoom on IC component.]{
    \includegraphics[width=0.48\textwidth]{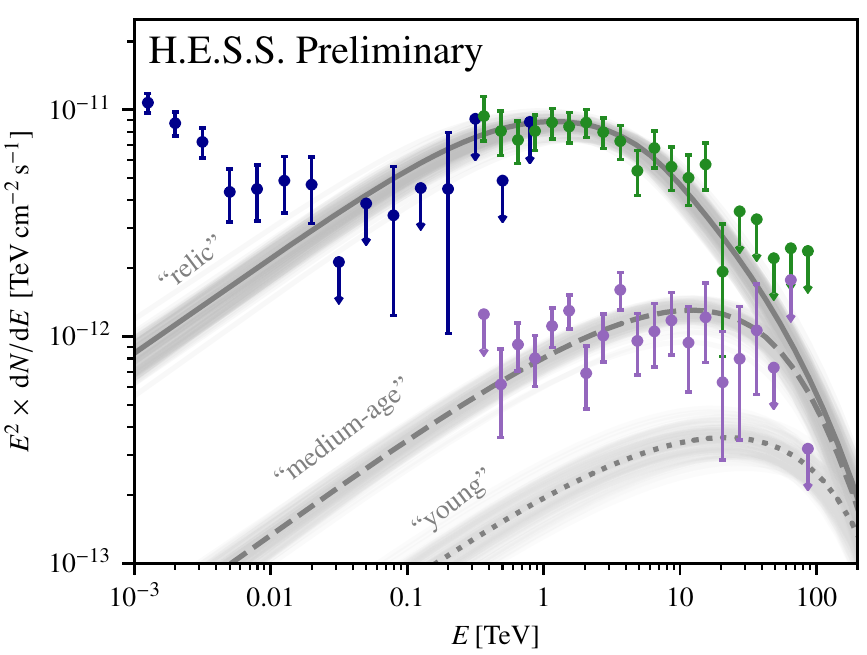}
    \label{fig:gamera_zoom}
  }
  \caption{
    SED of \hessj, with results of the PWN model.
    The thick lines display the best-fit model curves, whereas the thin lines display individual solutions of the MCMC sampling.
    The Suzaku data are from \cite{Anada2010} and the radio data for \snrS (not used in the fit) are from \cite{Brogan2006}.
  }
  \label{fig:gamera}
\end{figure}

First, we have modelled the entire emission of \hessj in a leptonic (PWN) scenario.
We performed a time-dependent modelling that takes into account the pulsar braking, employing the GAMERA library \cite{Hahn2016}.
Two describe both \hess components and the X-ray nebula (which is offset from the peak in \gam-ray emission, cf.\ Fig.~\ref{fig:flux_maps}), we invoked three ``generations'' of electrons:\\
(i) ``relic'' electrons, associated with component~A and injected over the system life time ($\approx \SI{33}{kyr}$);\\
(ii) ``medium-age'' electrons, associated with component~B and injected within the last $\approx \SI{4.7}{kyr}$;\\
(iii) ``young'' electrons, associated with the X-ray nebula and injected within the last $\approx \SI{1.2}{kyr}$.\\
The results of the model are displayed in Fig.~\ref{fig:gamera}.
From the approximate age of the system and the measured extent of component~A, it is possible to derive a diffusion coefficient for the ``relic'' electrons.
We obtained $D\approx \SI{1e28}{\square\centi\meter\per\second}$, which is of the same order as the coefficient measured in the vicinity of the Geminga PWN \cite{HAWC2017}.
In such a scenario, one would furthermore expect a cut-off in the energy spectrum of component~A, as the highest-energy electrons should have cooled due to IC scattering by now.
This is consistent with the measured cut-off for this component at $\approx\SI{13}{TeV}$.
In summary, the PWN model shows that the \gam-ray emission of \hessj can be modelled in a PWN scenario, and that in particular component~A of \hessj can be well described as a halo of old electrons that surround the compact PWN.

The presence of SNRs and molecular clouds in the region motivates to also consider a hadronic scenario in which (part of) the emission is due to cosmic-ray nuclei accelerated by the SNRs and interacting with gas in the clouds.
We focus here in particular on component~B of \hessj, which coincides in position with the edge of \snrS and several of the dense molecular clouds (cf.\ Fig.~\ref{fig:flux_maps}).
Using the \texttt{Naima} package \cite{Zabalza2016}, we have fitted a proton-proton model to component~B, obtaining a required energy in primary protons of $W_p\sim\num{4e49}(n/\SI{1}{\per\cubic\centi\meter})^{-1}\,\mathrm{erg}$.
Considering that gas densities $\gg$\SI{1}{\per\cubic\centi\meter} are expected in the clouds \cite{Castelletti2016}, this presents a viable alternative interpretation.

\vspace{-0.3cm}
\section{Conclusion}\vspace{-0.3cm}
We have presented a new \hess analysis of the unassociated \gam-ray source \hessj.
For the first time, we were able to resolve the emission into two components that exhibit distinct spectra and morphologies.
Our \fermilat analysis has confirmed the presence of extended emission also in the GeV energy range, which is however challenging to associate with either of the components of \hessj.
The extended component~A of \hessj is compatible with a halo of old electrons around the compact PWN.
The compact component~B could plausibly be of either leptonic or hadronic origin.

\bibliographystyle{JHEP}
\bibliography{hessj1809}

\end{document}